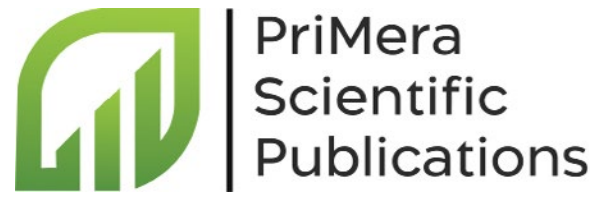

# Extending the (Elementary) Mathematical Data Model and *MatBase* with Two New Constraint Types: Inexistence and Anti-existence

**Christian Mancas***

*Mathematics and Computer Science Department, Ovidius University at Constanta, Romania*

***Corresponding Author:** Christian Mancas, Ovidius University, Bd. Mamaia 124, Constanta, CT, Romania.

## Abstract

This research paper introduces two new constraint types and four subtypes of database constraints added to our (Elementary) Mathematical Data Model, which are the duals of the existence and non-existence ones. They are formally defined, characterized, and exemplified with real-life instances. The well-formedness, satisfiability, coherence, and minimality of sets of all 7 subtypes of existence constraints are studied. Corresponding SQL-embedded pseudocode algorithms for managing such sets are provided and proved to be of constant complexity, sound, complete, and optimal. Also provided are algorithms for enforcing these new types of constraints, called inexistence and anti-existence. Their characterization proves that they have linear complexity in the sum of the arities of the function (Cartesian products) involved, and are sound, complete, and optimal as well. All these algorithms were implemented in both versions of our intelligent data and knowledge base management system prototype *MatBase*, which automatically generates code for enforcing all the 7 subtypes of existence constraints.

## Introduction

In [1] we introduced the non-existence constraints, in the framework of our (Elementary) Mathematical Data Model ((E)MDM) [2]. Recent conceptual data modeling projects brought to our attention examples of business rules whose formalization are duals to the existence and non-existence ones.

As all business rules governing subuniverses of interest must be formalized as constraints and enforced either by the underlying database (db) management systems (DBMSes) or/and by the software applications built ontop of them to guarantee data plausibility [1], the highest possible quality standard in dbs, we decided to incorporate these new constraint types into (E)MDM, under the names of inexistence and anti-existence constraints, respectively.

As, just like for the non-existence ones, each of them has two subtypes, namely single and consolidated ones, this extension brings the constraint types total number of the (E)MDM to 80. Of course, we added them too to both versions of *MatBase* (the MS Access one for small and medium dbs and the C# and SQL Server one for large dbs), our intelligent data and knowledge base management system prototype [3] based on both (E)MDM, Entity-Relationship Data Model (RDM) [4-6], and Relational Data Model [6-8].

The next Section mentions related work. The third one is dedicated to the materials and methods used. The fourth one presents and discusses the results obtained. The paper ends with conclusions and a list of references.

## Related Work

Among the famous 12 rules of E.F. Codd, the father of the RDM, the third one refers to null values [9]: "*Null values (distinct from the empty character string or a string of blank characters and distinct from zero or any other number) are supported in fully relational DBMS for representing missing information and inapplicable information in a systematic way, independent of data type*". Codd also introduced the null substitution principle [10].

Dozens of papers were dedicated to the theoretical and practical treatment of null values in RDM and relational DBMSes (see, e.g., [11]).

Existence constraints were introduced in [12]. We extended them to composite functions and function products [1, 2].

A first algorithm on enforcing coherence and minimality of (E)MDM constraint sets was published in [13].

## Materials and Methods

Recall that a *function* (*mapping*) is a binary relation between two sets, not necessarily distinct, that satisfies two conditions: *functionality*, i.e., each element of one set (called the *domain*) is mapped to only one element of the other set (called the *codomain*), and *totality*, i.e., any element of the domain is mapped to one element of the codomain; obviously, they can be combined into only on condition: each element from the domain is mapped to exactly one element from the codomain.

Using the standard math notation, e.g., *Country* : *CITIES* → *COUNTRIES* is a function (between the domain *CITIES* and the codomain *COUNTRIES*) as it maps any world city to the country to which it belongs (where territories which are not countries, like Gaza, are assimilated in *COUNTRIES* as well). For example, *Country*(New York) = U.S.A., *Country*(Montreal) = Canada, *Country*(Sydney) = Australia, *Country*(Sibiu) = Romania, etc.

Often, not only in dbs, we either temporarily do not know or do not care what are the values of a function for some elements of its domain, or they even do not exist. For example, *Capital* : *COUNTRIES* → *CITIES* do not have values not only for some territories but even for the island country of Nauru; moreover, in relational dbs (rdbs), due to the circular nature of these two functions, you have to first add to the table *COUNTRIES* a new entry (say, for France) and only then add to the table *CITIES* entries for its cities (say, Paris, its capital, Marseille, Strasbourg, etc.). This means that *Capital* must not be totally defined, or *accept null values*, or be *partially defined*, which, theoretically, means that its codomain is extended to *CITIES* ∪ *NULLS*, where NULLS is a distinguished countable set of *null values* (where *distinguished* means that none of its elements also belong to another set and *countable* means that is discretely infinite, just like the set of naturals).

Consequently, in general, by default, all rdbs table columns (i.e., functions defined on the sets of rows making the instances of the corresponding tables and taking values either from such instances, if they are foreign keys, or from data types, i.e., subsets of naturals, rationals, reals, string sets over the ASCII or UNICODE alphabets, etc.) are not totally defined (i.e., they accept null values) and totality is an explicit constraint (called NOT NULL) associated with some of them.

Mathematically, the schema of a rdb table is a function (Cartesian) product, e.g., *COUNTRIES* is a function product *x* · *CountryName* · *Capital* · *Currency* · *Population* · *Area* : *COUNTRIES* → NAT(3) × ASCII(255) × *CITIES* × *CURRENCIES* × [450, 10000000000] × [0.4, 20000000], where NAT(3) = {0, 1, …, 999}, ASCII(255) is the subset of strings made of ASCII characters and having maximum length 255, x is a surrogate autonumber primary key, *CountryName* is NOT NULL and unique (i.e., one-to-one, as there may not be two countries or territories having same name), *Capital* is unique as well (as no city may simultaneously be the capital of two countries), and *Capital* and *Currency* are foreign keys referencing *CITIES* and *CURRENCIES*, respectively (hopefully, pointing to their corresponding primary keys *x*).

Generally, given functions $f : D \to C_f$ and $g : D \to C_g$, their *function* (*Cartesian*) *product* is the function denoted $f \cdot g : D \to C_f \times C_g$, with $(f \cdot g)(x) = <f(x), g(x)>$, for any element $x$ of $D$, where × is the symbol denoting the *Cartesian product set* operator: e.g., if $C_f = \{1,2\}$ and $C_g = \{a,b\}$, then $C_f \times C_g = \{<1,a>, <1,b>, <2,a>, <2,b>\}$. *f* and *g* are called *single*.

Mathematically, SQL JOIN operators perform function compositions; e.g., to obtain the set of pairs <country name, capital name> you need to run the query SELECT CountryName, CityName FROM COUNTRIES JOIN CITIES ON COUNTRIES.Capital = CITIES.x, which computes the function product *CountryName* · *Capital* ° *CityName*, where *Capital* ° *CityName* : *COUNTRIES* → ASCII(255) is the *composition* of *Capital* with *CityName* : CITIES → ASCII(255).

Generally, given functions $f : D \to C$ and $g : C \to E$, the *composite function* $f \circ g : D \to E$ computes $(f \circ g)(x) = f(g(x))$, for any element $x$ of $D$. For composite functions, *f* and *g* are called *atomic*. In conceptual data modeling, only products including composite functions are of interest, while their dual, i.e., compositions including function products, are not.

Conceptual data modeling, db design, and software engineering must always rigorously treat null values, as, otherwise, either implausible data is stored in dbs, or incorrect results are computed by queries, or both. Consequently, (E)MDM included for many years already the existence and non-existence constraint types.

Recall that an *existence constraint* is written $f \mid\!- g$ and is a shorthand for $(\forall x \in D)(f(x) \notin \text{NULLS} \Rightarrow g(x) \notin \text{NULLS})$, i.e., anytime when *f* is defined, *g* must be defined as well, e.g., *email* |– *FirstName* · *LastName*, for the business rule “Whenever the email address of somebody is known, the corresponding first and last names must also be known”. Generally, to the left of the implication operator |– there may also be a function product: $f_1 \cdot \ldots \cdot f_n \mid\!- g_1 \cdot \ldots \cdot g_m$, $n, m > 0$, naturals, with both $f_1 \cdot \ldots \cdot f_n$ and $g_1 \cdot \ldots \cdot g_m$ being defined over the same set $D$, is a shorthand for $(\forall x \in D)(\exists i, 1 \le i \le n)(f_i(x) \notin \text{NULLS} \Rightarrow g_j(x) \notin \text{NULLS}, \forall j, 1 \le j \le m)$, i.e., whenever at least one $f_i$ ‘s value is known, all $g_j$ ‘s values must also be known, e.g., *streetAddress* · *email* |– *FirstName* · *LastName* is formalizing the business rule “Whenever either the street or email addresses of somebody is/are known, the corresponding first and last names must also be known”.

Right-dually, we introduced the concept of *non-existence constraints*, written $f \neg\mid\!- g$, which is a shorthand for $(\forall x \in D)(f(x) \notin \text{NULLS} \Rightarrow g(x) \in \text{NULLS})$, i.e., anytime when *f* is defined, *g* must not be defined, e.g., *SSN* ¬|– *ITIN*, for the business rule “Whenever a U.S. resident has a SSN, then he/she may not also have an ITIN” (where SSN = Social Security Number and ITIN = Individual Taxpayer Identification Number) [1]. As in the majority of cases these constraints are symmetric, e.g., *ITIN* ¬|– *SSN* also holds, we also introduced the simpler, coalescing syntax $\neg\mid\!- f_1 \cdot \ldots \cdot f_n$, which stands for $(\forall x \in D)(\exists i, 1 \le i \le n)(f_i(x) \notin \text{NULLS} \Rightarrow f_j(x) \in \text{NULLS}, \forall j, 1 \le j \le n, j \ne i)$, i.e., whenever one $f_i$ ‘s value is known, all other $f_j$ ‘s values must be unknown (alternatively, at most one $f_i$ ‘s value may be known for any element of $D$), e.g., ¬|– *SSN* · *ITIN* is formalizing the business rule “U.S. citizens may have assigned either a SSN or an ITIN but not both”.

The first new existence-type constraint we are adding to (E)MDM in this paper is the dual of the non-existence: an *inexistence constraint* is written as $\neg f \mid\!- g$, which abbreviates $(\forall x \in D)(f(x) \in \text{NULLS} \Rightarrow g(x) \notin \text{NULLS})$, i.e., anytime when *f* is unknown, *g* must be known, e.g., *email* ¬|– *PhoneNo*, for the business rule “Whenever the email address of a contact is unknown, the corresponding phone number must be known as well”. As in the majority of cases these constraints are symmetric too, e.g., *PhoneNo* ¬|– *email* also holds, we also introduced the simpler, coalescing syntax $\mid\!- f_1 \cdot \ldots \cdot f_n$, which stands for $(\forall x \in D)(\exists i, 1 \le i \le n)(f_i(x) \notin \text{NULLS})$, i.e., at least one $f_i$ ‘s value must be known for any element of $D$, e.g., ¬|– *email* · *PhoneNo* is formalizing the business rule “For every contact, either the email address or the phone number must be known”. Please do not confuse this (E)MDM notation with the similar RDM one, which means “all $f_i$ ‘s are NOT NULL”: recall that *f* NOT NULL from RDM is written *f* total in (E)MDM.

Finally, the second new existence-type constraint we are adding to (E)MDM in this paper is the dual of the existence one: an *anti-existence* constraint is written as $\neg f \neg\vdash g$, which abbreviates $(\forall x \in D)(f(x) \in \text{NULLS} \Rightarrow g(x) \in \text{NULLS})$, i.e., anytime when $f$ is unknown, $g$ must also be unknown, e.g., $\neg$*PassedAwayYear* $\neg\vdash$ *KilledBy*, for the “business” rule “Whenever the passed away year of a person is unknown, that person was not killed” (famously put it as “There’s no living with a killing” in the iconic *Shane* Western). We also introduced the simpler, coalescing syntax $\neg\neg\vdash f_1 \cdot \ldots \cdot f_n$, which stands for $(\forall x \in D)(\exists i, 1 \leq i \leq n)(f_i(x) \in \text{NULLS} \Rightarrow f_j(x) \in \text{NULLS}, \forall j, 1 \leq j \leq n)$, i.e., all $f_i$ ‘s values are either known or unknown, for any element of $D$, e.g., $\neg\neg\vdash$ *MultipleOf* $\cdot$ *MultiplicityFactor* is formalizing the business rule “Any measurement unit is either a multiple of another one, case in which the corresponding multiplicity factor must also be known, or it is not, case in which the multiplicity factor does not make sense, hence it must also be unknown.” Indeed, e.g., the MB is a multiple of the KB with factor 1024 but the bit is not multiple of anything, hence it would be senseless to associate a multiplicity factor with it. Proposition 7 proves that $\neg\neg\vdash f \cdot g$ is equivalent to $\{f \vdash g, \neg f \neg\vdash g\}$.

### *MatBase implementation*

To accommodate storing anti-existence and inexistence constraints together with the existence and non-existence ones, we simply added to *MatBase’s* metacatalog table *EXIST_CNSTRS* [1] another Boolean column called *Inexist*?. Figure 1 shows its augmented schema and a fragment of its instance.

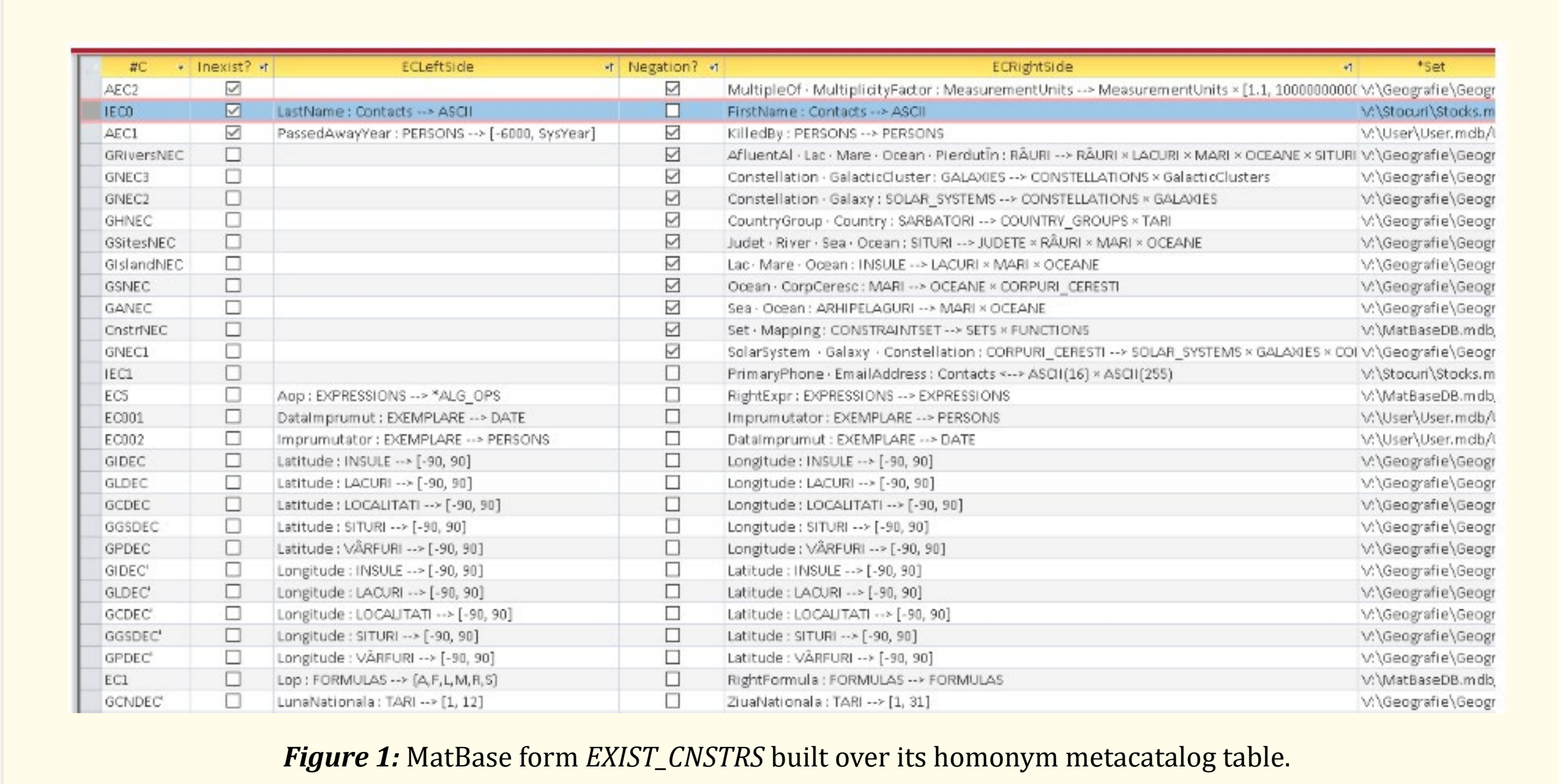

| #C | Inexist? | ECLeftSide | Negation? | ECRightSide | *Set |
|---|---|---|---|---|---|
| AEC2 | ☑ | | ☑ | MultipleOf · MultiplicityFactor : MeasurementUnits --> MeasurementUnits × [1.1, 10000000000( | V:\Geografie\Geogr |
| IEC0 | ☑ | LastName : Contacts --> ASCII | ☐ | FirstName : Contacts --> ASCII | V:\Stocuri\Stocks.m |
| AEC1 | ☑ | PassedAwayYear : PERSONS --> [-6000, SysYear] | ☑ | KilledBy : PERSONS --> PERSONS | V:\User\User.mdb/\ |
| GRiversNEC | ☐ | | ☑ | AfluentAl · Lac · Mare · Ocean · Pierdutîn : RÂURI --> RÂURI × LACURI × MARI × OCEANE × SITURI | V:\Geografie\Geogr |
| GNEC3 | ☐ | | ☑ | Constellation · GalacticCluster : GALAXIES --> CONSTELLATIONS × GalacticClusters | V:\Geografie\Geogr |
| GNEC2 | ☐ | | ☑ | Constellation · Galaxy : SOLAR_SYSTEMS --> CONSTELLATIONS × GALAXIES | V:\Geografie\Geogr |
| GHNEC | ☐ | | ☑ | CountryGroup · Country : SARBATORI --> COUNTRY_GROUPS × TARI | V:\Geografie\Geogr |
| GSitesNEC | ☐ | | ☑ | Judet · River · Sea · Ocean : SITURI --> JUDETE × RÂURI × MARI × OCEANE | V:\Geografie\Geogr |
| GIslandNEC | ☐ | | ☑ | Lac · Mare · Ocean : INSULE --> LACURI × MARI × OCEANE | V:\Geografie\Geogr |
| GSNEC | ☐ | | ☑ | Ocean · CorpCeresc : MARI --> OCEANE × CORPURI_CERESTI | V:\Geografie\Geogr |
| GANEC | ☐ | | ☑ | Sea · Ocean : ARHIPELAGURI --> MARI × OCEANE | V:\Geografie\Geogr |
| CnstrNEC | ☐ | | ☑ | Set · Mapping : CONSTRAINTSET --> SETS × FUNCTIONS | V:\MatBaseDB.mdb, |
| GNEC1 | ☐ | | ☑ | SolarSystem · Galaxy · Constellation : CORPURI_CERESTI --> SOLAR_SYSTEMS × GALAXIES × COI | V:\Geografie\Geogr |
| IEC1 | ☐ | | ☐ | PrimaryPhone · EmailAddress : Contacts <--> ASCII(16) × ASCII(255) | V:\Stocuri\Stocks.m |
| EC5 | ☐ | Aop : EXPRESSIONS --> *ALG_OPS | ☐ | RightExpr : EXPRESSIONS --> EXPRESSIONS | V:\MatBaseDB.mdb, |
| EC001 | ☐ | DataImprumut : EXEMPLARE --> DATE | ☐ | Imprumutator : EXEMPLARE --> PERSONS | V:\User\User.mdb/\ |
| EC002 | ☐ | Imprumutator : EXEMPLARE --> PERSONS | ☐ | DataImprumut : EXEMPLARE --> DATE | V:\User\User.mdb/\ |
| GIDEC | ☐ | Latitude : INSULE --> [-90, 90] | ☐ | Longitude : INSULE --> [-90, 90] | V:\Geografie\Geogr |
| GLDEC | ☐ | Latitude : LACURI --> [-90, 90] | ☐ | Longitude : LACURI --> [-90, 90] | V:\Geografie\Geogr |
| GCDEC | ☐ | Latitude : LOCALITATI --> [-90, 90] | ☐ | Longitude : LOCALITATI --> [-90, 90] | V:\Geografie\Geogr |
| GGSDEC | ☐ | Latitude : SITURI --> [-90, 90] | ☐ | Longitude : SITURI --> [-90, 90] | V:\Geografie\Geogr |
| GPDEC | ☐ | Latitude : VÂRFURI --> [-90, 90] | ☐ | Longitude : VÂRFURI --> [-90, 90] | V:\Geografie\Geogr |
| GIDEC' | ☐ | Longitude : INSULE --> [-90, 90] | ☐ | Latitude : INSULE --> [-90, 90] | V:\Geografie\Geogr |
| GLDEC' | ☐ | Longitude : LACURI --> [-90, 90] | ☐ | Latitude : LACURI --> [-90, 90] | V:\Geografie\Geogr |
| GCDEC' | ☐ | Longitude : LOCALITATI --> [-90, 90] | ☐ | Latitude : LOCALITATI --> [-90, 90] | V:\Geografie\Geogr |
| GGSDEC' | ☐ | Longitude : SITURI --> [-90, 90] | ☐ | Latitude : SITURI --> [-90, 90] | V:\Geografie\Geogr |
| GPDEC' | ☐ | Longitude : VÂRFURI --> [-90, 90] | ☐ | Latitude : VÂRFURI --> [-90, 90] | V:\Geografie\Geogr |
| EC1 | ☐ | Lop : FORMULAS --> {A,F,L,M,R,S} | ☐ | RightFormula : FORMULAS --> FORMULAS | V:\MatBaseDB.mdb, |
| GCNDEC' | ☐ | LunaNationala : TARI --> [1, 12] | ☐ | ZiuaNationala : TARI --> [1, 31] | V:\Geografie\Geogr |

***Figure 1:*** MatBase form *EXIST_CNSTRS* built over its homonym metacatalog table.

*#C* is the primary key of this table and a foreign key referencing the primary key of the metacatalog table *CONSTRAINTSET*, which is storing for all constraints known to *MatBase* their names, dbs, type, logic formula, semantics (with database · formula, database · name, and database · semantics unique keys), etc. Mathematically, *#C* is the canonical inclusion injection associated to the set inclusion *EXIST_CNSTRS* ⊆ *CONSTRAINTSET* (i.e., existence constraints are constraints). *MatBase* shows constraint names instead of their automatically generated integer identification values. *ECLeftSide* and *ECRightSide* show the functions that make the left- and right-side, respectively, of these constraints (hence, you can imagine that the implication operator |– is stored between *Negation*? and *ECRightSide*). *Inexist*? stores the negation (¬) operator (if any) before *ECLeftSide*, while *Negation*? stores the one before the implication operator |–. Finally, the computed column **Set* stores the common domain set of *ECLeftSide* and *ECRightSide*; e.g., for the inexistence constraint IEC0: ¬LastName |– FirstName (see second, selected line of data from Figure 1), formalizing the “business” rule “Whenever the last name of somebody is not known, the corresponding first name must be known” (which must be enforced not only for companies, which generally have names of only one word, but also for humans: e.g., ancient Greeks did not use last names, see Homer, Sophocles, Aristotle, etc.), the common domain set is *Contacts* from the db *Stocks.mdb* (stored in the *Stocuri* folder of the virtual logic drive V:).

The values of *Inexist*? and *Negation*?, as well as the presence or absence of *ECLeftSide* determine the subtype of existence constraints, as shown in Figure 2 (where 0 = *False* and 1 = *True*).

| *Inexist?* | *Negation?* | *Constraint Type* | *Syntax* | *Formula* |
|---|---|---|---|---|
| 0 | 0 | Existence | $f \vdash g$ | $f(x) \notin \text{NULLS} \Rightarrow g(x) \notin \text{NULLS}$ |
| 0 | 1 | Non-Existence | $f \neg\vdash g$<br>$\neg\vdash f \bullet g$ | $f(x) \notin \text{NULLS} \Rightarrow g(x) \in \text{NULLS}$<br>$f(x) \in \text{NULLS} \wedge g(x) \notin \text{NULLS} \vee$<br>$f(x) \notin \text{NULLS} \wedge g(x) \in \text{NULLS} \vee$<br>$f(x) \in \text{NULLS} \wedge g(x) \in \text{NULLS}$ |
| 1 | 0 | Inexistence | $\neg f \vdash g$<br>$\vdash f \bullet g$ | $f(x) \in \text{NULLS} \Rightarrow g(x) \notin \text{NULLS}$<br>$f(x) \in \text{NULLS} \wedge g(x) \notin \text{NULLS} \vee$<br>$f(x) \notin \text{NULLS} \wedge g(x) \in \text{NULLS} \vee$<br>$f(x) \notin \text{NULLS} \wedge g(x) \notin \text{NULLS}$ |
| 1 | 1 | Anti-Existence | $\neg f \neg\vdash g$<br>$\neg\neg\vdash f \bullet g$ | $f(x) \in \text{NULLS} \Rightarrow g(x) \in \text{NULLS}$<br>$f(x) \notin \text{NULLS} \wedge g(x) \notin \text{NULLS} \vee$<br>$f(x) \in \text{NULLS} \wedge g(x) \in \text{NULLS}$ |

***Figure 2:*** The seven existence constraint subtypes.

Please note that existence and inexistence are left-duals, just like non-existence and anti-existence, while existence and non-existence are right-duals, just like inexistence and anti-existence.

***Well-formed existence constraints***

The existence constraints of no matter what subtype are *well-formed* iff they have the syntaxes shown in Figure 2 and, moreover, obey the following Propositions 0 to 5:

*Proposition* 0. Let *ec* be any existence constraint of no matter what subtype;

i. Both *ECLeftSide* and *ECRightSide* must be defined over the same domain set.
ii. *ECRightSide* : *EXIST_CNSTRS* → *FUNCTIONS* must be totally defined.

*Proof*: Trivial, as:

i. otherwise, they may not be members of the same function product;
ii. no well-formed ec exists in Figure 2 without an *ECRightSide* (see column *Syntax*). *Q.E.D.*

*Corollary* 0. *MatBase* must reject any (anti/non/in/)existence constraint without right-hand side or with sides that are not defined over a same set.

*Proposition* 1. All single functions from both *ECLeftSide* and *ECRightSide* must be partially defined.

*Proof*: Trivial, as otherwise they would not take null values. *Q.E.D.*

*Corollary* 1. Let *ec* be any existence constraint of no matter what subtype;

i. Any composite function taking part in *ec* must contain at least one partially defined atomic function.
ii. The only totally defined functions that may appear in *ec* must be atomic.
iii. *MatBase* must reject any (anti/non/in/)existence constraint involving a totally defined function, except when that function is a member of a partially defined composite one.

*Proof*: Trivial, left to the reader (hint, e.g., *Capital* ° *CityName* above is not total, although *CityName* is total but *Capital* is not). *Q.E.D.*

(E)MDM and *MatBase* consider both a composite and a (Cartesian) product function to be partially defined whenever at least one of their member functions is not totally defined (see [2] and [1], respectively).

*Proposition* 2. No atomic function may appear more than once in an existence constraint of no matter what subtype.

*Proof*: Obviously, as on the same side it would be superfluous and on both sides it would be either superfluous or incoherent. *Q.E.D.*

Please note that (E)MDM and *MatBase* do not accept duplicates in function products (e.g., see the constraint *uniq_ FUNCT_PRODUCTS_F* from VIII.D of [14]).

*Corollary* 2. In any existence constraint of no matter what subtype,

i. *MatBase* must enforce $\{f_1, \ldots, f_n\} \cap \{g_1, \ldots, g_m\} = \emptyset$
ii. *ECLeftSide* ≠ *ECRightSide*.

*Proof*: Trivial, left to the reader.

***Coherence and minimality of the sets of (anti/non/in/)existence constraints***

*Proposition* 3. To guarantee that *EXIST_CNSTRS* is a set and that any set of existence constraints is minimal, *ECSTS* = *ECRightSide* · *ECLeftSide* · *Inexist*? · *Negation*? : *EXIST_CNSTRS* ↔ *FUNCTIONS* × (*FUNCTIONS* ∪ *NULLS*) × BOOLE × BOOLE must be minimally injective (one-to-one).

*Proof*: Let us denote $f = 1$, $g = 2$, $f \cdot g = 3$, absence of $f$ and *False* with 0, and *True* = 1; column *Values* from Figure 3 shows the corresponding values for all 7 existence constraint subtypes from Figure 2, obtained by concatenating the values of these 4 columns (presented in the order in which they are stored in both the table and the form from Figure 1): {13, 102, 112, 1002, 1013, 1102, 1112}. As expected, there are no duplicates, so *ECSTS* is injective. Column *MinValues* shows the corresponding values when eliminating *ECLeftSide*: {2, 12, 13, 102, 103, 112, 113}; the fact that there are no duplicates either, means that, syntactically, you can uniquely identify existence constraint subtypes even without considering *ECLeftSide*. However, to also eliminate duplicates from any existence constraint set, *ECLeftSide* must be part of this key; otherwise, e.g., users might store in a same db, under distinct names, two constraints of the types $f$ |– $g$ or $f$ ¬|– $g$ having the same body. Moreover, if *Inexist*? is eliminated from *ECSTS*, 2 and 12 are duplicated; if *Negation*? is eliminated, 2 and 12 are duplicated; finally, if *ECRightSide* is eliminated, 1, 10, and 11 are duplicated. Consequently, *ECSTS* is minimally injective. *Q.E.D.*

*Corollary* 3. To avoid storing duplicates in the *EXIST_CNSTRS* table, *MatBase* must enforce the unique semantic key *ECRightSide* · *ECLeftSide* · *Inexist*? · *Negation*?

| *No.* | *Syntax* | *Inexist?* | *ECLeftSide* | *Negation?* | *ECRightSide* | *Value* | *MinValue* |
|---|---|---|---|---|---|---|---|
| 1 | $f$ \|— $g$ | 0 | 1 | 0 | 2 | 0102 | 2 |
| 2 | $f$ ¬\|— $g$ | 0 | 1 | 1 | 2 | 0112 | 12 |
| 3 | ¬\|— $f$ • $g$ | 0 | 0 | 1 | 3 | 0013 | 13 |
| 4 | ¬$f$ \|— $g$ | 1 | 1 | 0 | 2 | 1102 | 102 |
| 5 | \|— $f$ • $g$ | 1 | 0 | 0 | 3 | 1003 | 103 |
| 6 | ¬$f$ ¬\|— $g$ | 1 | 1 | 1 | 2 | 1112 | 112 |
| 7 | ¬¬\|— $f$ • $g$ | 1 | 0 | 1 | 3 | 1013 | 113 |

***Figure 3:*** The seven existence constraint subtypes syntax abbreviated, with and w/o *ECLeftSide*.

*Proposition* 4. In any *ec*, whenever *ECLeftSide* is not specified, then *ECRightSide* must be a function product (i.e., $*arity(ECLeftSide) = 0 \Rightarrow *arity(ECRightSide) > 1$).

*Proof*: trivial, from inspecting the table from Figure 3: *ECLeftSide* is missing only for subtypes 3, 5, and 7, for which *ECRightSide* is always $f \cdot g$. *Q.E.D.*

*Corollary* 4. *MatBase* must enforce $*arity(ECLeftSide) = 0 \Rightarrow *arity(ECRightSide) > 1$, for any compacted anti- or in-existence constraint.

*Proposition* 5. Considering False equivalent to a null value (as, usually in dbs, whenever a Boolean column is left null, that null is considered *False*):

i. $\neg(Inexist? \cdot Negation?) \mid\!- ECLeftSide$ (any not negated not inexistence constraint is an existence one, hence *ECLeftSide* is compulsory).
ii. $\neg(Inexist? \cdot ECLeftSide) \mid\!- Negation?$ (any not inexistence constraint without left side is a non-existence one, hence *Negation*? must be *True*).

*Proof*:

i. Trivially, if we assume by *reductio ad absurdum* that *ECLeftSide* may be null in such circumstances, it follows that existence constraints having syntax $\mid\!- g$ with *Inexist*? = *False* are valid, which is absurd.
ii. Similarly, if we assume by *reductio ad absurdum* that *Negation*? may be null in such circumstances, it follows that existence constraints having syntax $\mid\!- g$ with *Inexist*? = *False* are valid, which is absurd. *Q.E.D.*

*Corollary* 5. *MatBase* must reject any existence constraint without left-hand side, as well as any non-existence one without *Negation*? = *True*.

The seven existence constraint subtypes also interact between them, not only with other (E)MDM constraint types. To simplify proofs of the following results, please note the following:

- As $f$ and $g$ are interchangeable, $(\forall x \in D)(f(x) \in NULLS \Rightarrow g(x) \notin NULLS)$ is equivalent to $(\forall x \in D)(f(x) \in NULLS \Rightarrow g(x) \notin NULLS \vee g(x) \in NULLS \Rightarrow f(x) \notin NULLS)$ and $(\forall x \in D)(f(x) \notin NULLS \Rightarrow g(x) \in NULLS)$ is equivalent to $(\forall x \in D)(f(x) \notin NULLS \Rightarrow g(x) \in NULLS \vee g(x) \notin NULLS \Rightarrow f(x) \in NULLS)$. Moreover, as $a \Rightarrow b$ is equivalent to $\neg a \vee b$, these formulas are equivalent to $(\forall x \in D)(f(x) \notin NULLS \vee g(x) \notin NULLS \vee g(x) \notin NULLS \vee f(x) \notin NULLS)$, i.e., $(\forall x \in D)(f(x) \notin NULLS \vee g(x) \notin NULLS)$, and $(\forall x \in D)(f(x) \in NULLS \vee g(x) \in NULLS \vee g(x) \in NULLS \vee f(x) \in NULLS)$, i.e., $(\forall x \in D)(f(x) \in NULLS \vee g(x) \in NULLS)$, respectively.
- Although all formulas from column *Formula* of the table from Figure 2 are first-order predicate logic ones, they can be computed as their propositional (zero-order) logic equivalent one, easily obtainable, e.g., by denoting $f(x) \in NULLS = f$ and $g(x) \in NULLS = g$; e.g., the two above ones become $(f \Rightarrow \neg g) \vee (g \Rightarrow \neg f)$, which is, indeed, equivalent to $\neg f \vee \neg g$, and $(\neg f \Rightarrow g) \vee (\neg g \Rightarrow f)$, which is equivalent to $f \vee g$, respectively. Figure 4 provides the propositional logic formulas for all existence constraint subtypes from Figure 2.

According to Figure 4, in fact, only 5 subtypes are independent of each other: being trivial, we leave proof of Proposition 6 to the reader.

*Proposition* 6.

i. $f \neg\mid\!- g \Leftrightarrow \neg\mid\!- f \cdot g$.
ii. $\neg f \mid\!- g \Leftrightarrow \mid\!- f \cdot g$.

| *Constraint Type* | *Syntax* | *Formula* | *Propositional Formula* |
|---|---|---|---|
| Existence | $f \,|\!-\, g$ | $f(x) \notin NULLS \Rightarrow g(x) \notin NULLS$ | $f \vee \neg g$ |
| Non-Existence | $f \neg|\!-\, g$<br>$\neg|\!-\, f \bullet g$ | $f(x) \notin NULLS \Rightarrow g(x) \in NULLS$<br>$f(x) \in NULLS \wedge g(x) \notin NULLS \vee f(x) \notin NULLS \wedge g(x) \in NULLS$<br>$\vee f(x) \in NULLS \wedge g(x) \in NULLS$ | $f \vee g$<br>$f \vee g$ |
| Inexistence | $\neg f \,|\!-\, g$<br>$|\!-\, f \bullet g$ | $f(x) \in NULLS \Rightarrow g(x) \notin NULLS$<br>$f(x) \in NULLS \wedge g(x) \notin NULLS \vee$<br>$f(x) \notin NULLS \wedge g(x) \in NULLS \vee f(x) \notin NULLS \wedge g(x) \notin NULLS$ | $\neg f \vee \neg g$<br>$\neg f \vee \neg g$ |
| Anti-Existence | $\neg f \neg|\!-\, g$<br>$\neg\neg|\!-\, f \bullet g$ | $f(x) \in NULLS \Rightarrow g(x) \in NULLS$<br>$f(x) \notin NULLS \wedge g(x) \notin NULLS \vee f(x) \in NULLS \wedge g(x) \in NULLS$ | $\neg f \vee g$<br>$f \leftrightarrow g$ ($f$ XNOR $g$) |

***Figure 4:*** The seven existence constraint subtypes and their corresponding logic propositions.

*Corollary* 6. According to Proposition 6 and Figure 4, *MatBase* must always replace:

i. $f \neg|\!- g$ by $\neg|\!- f \cdot g$ (as it is explicitly richer, also including $f(x) \in NULLS \wedge g(x) \in NULLS$).
ii. $\neg f \,|\!- g$ by $|\!- f \cdot g$ (as it is explicitly richer, also including $f(x) \notin NULLS \wedge g(x) \notin NULLS$).

Analyzing the interactions between the C(5, 2) = 10 pairs of the remaining independent subtypes, we get the following results:

*Proposition* 7.

i. $f \,|\!- g \wedge \neg f \neg|\!- g \Leftrightarrow \neg\neg|\!- f \cdot g$.
ii. $f \,|\!- g \wedge \neg\neg|\!- f \cdot g \Leftrightarrow \neg\neg|\!- f \cdot g$.
iii. $\neg f \neg|\!- g \wedge \neg\neg|\!- f \cdot g \Leftrightarrow \neg\neg|\!- f \cdot g$.

*Proof*:

i. $f \,|\!- g \wedge \neg f \neg|\!- g$ can be computed as $(f \vee \neg g) \wedge (\neg f \vee g)$, which is equal to $f \leftrightarrow g$.
ii. $f \,|\!- g \wedge \neg\neg|\!- f \cdot g$ can be computed as $(f \vee \neg g) \wedge f \leftrightarrow g$, which is equal to $f \leftrightarrow g$.
iii. $\neg f \neg|\!- g \wedge \neg\neg|\!- f \cdot g$ can be computed as $(\neg f \vee g) \wedge f \leftrightarrow g$, which is equal to $f \leftrightarrow g$. *Q.E.D.*

*Corollary* 7. *MatBase* must always replace $f \,|\!- g \wedge \neg f \neg|\!- g$, $f \,|\!- g \wedge \neg\neg|\!- f \cdot g$, and $\neg f \neg|\!- g \wedge \neg\neg|\!- f \cdot g$ by $\neg\neg|\!- f \cdot g$.

*Proposition* 8. The following 6 existence constraint subtype pairs are mutually exclusive (i.e., orthogonal to each other):

$<f \,|\!- g, f \neg|\!- g>$, $<f \,|\!- g, \neg f \,|\!- g>$, $<f \neg|\!- g, \neg f \neg|\!- g>$, $<f \neg|\!- g, \neg\neg|\!- f \cdot g>$, $<\neg f \,|\!- g, \neg f \neg|\!- g>$, $<\neg f \,|\!- g, \neg\neg|\!- f \cdot g>$

*Proof*: Let us consider any $f : D \to C_f \cup NULLS$ and $g : D \to C_g \cup NULLS$;

i. (both $f \,|\!- g$ and $f \neg|\!- g$)

   Suppose by *reductio ad absurdum* that $f \,|\!- g \wedge f \neg|\!- g$, i.e., $(\forall x \in D)(f(x) \notin NULLS \Rightarrow g(x) \notin NULLS \wedge f(x) \notin NULLS \Rightarrow g(x) \in NULLS)$; trivially (by *tertium non datur*), $g(x)$ may not be both null and not null, for no $x$.

   Check through propositional calculus: $f \,|\!- g \wedge f \neg|\!- g$ can be computed as $(f \vee \neg g) \wedge (f \vee g)$, which is equal to $f$, i.e., $(\forall x \in D)(f(x) \in NULLS)$, which is absurd.

ii. (both $f \,|\!- g$ and $\neg f \,|\!- g$)

   Suppose by *reductio ad absurdum* that $f \,|\!- g \wedge \neg f \,|\!- g$, i.e., $(\forall x \in D)(f(x) \notin NULLS \Rightarrow g(x) \notin NULLS \wedge f(x) \in NULLS \Rightarrow g(x) \notin NULLS)$;

trivially, this would mean that $g(x)$ may never take null values, i.e., it is totally defined, which contradicts ¬($g$ total).

Check through propositional calculus: $f \,|\!-\, g \wedge \neg f \,|\!-\, g$ can be computed as $(\neg f \vee g) \wedge (f \vee g)$, which is equal to $g$, i.e., $(\forall x \in D)(g(x) \in \text{NULLS})$, which is absurd.

iii. (both $f \,\neg|\!-\, g$ and $\neg f \,\neg|\!-\, g$)

Suppose by *reductio ad absurdum* that $f \,\neg|\!-\, g \wedge \neg f \,\neg|\!-\, g$, i.e., $(\forall x \in D)(f(x) \notin \text{NULLS} \Rightarrow g(x) \in \text{NULLS} \wedge f(x) \in \text{NULLS} \Rightarrow g(x) \in \text{NULLS})$, for any $f$ and $g$; trivially, this is incoherent, as it would mean that $g$ takes only null values.

Check through propositional calculus: $f \,\neg|\!-\, g \wedge \neg f \,\neg|\!-\, g$ can be computed as $(f \vee g) \wedge (\neg f \vee g)$, which is equal to $g$, i.e., $(\forall x \in D)(g(x) \in \text{NULLS})$, which is absurd.

iv. (both $f \,\neg|\!-\, g$ and $\neg\neg|\!-\, f \cdot g$)

Suppose by *reductio ad absurdum* that $f \,\neg|\!-\, g \wedge \neg\neg|\!-\, f \cdot g$, i.e., $(\forall x \in D)(f(x) \notin \text{NULLS} \Rightarrow g(x) \in \text{NULLS} \wedge (f(x) \notin \text{NULLS} \wedge g(x) \notin \text{NULLS} \vee f(x) \in \text{NULLS} \wedge g(x) \in \text{NULLS}))$, which reduces to $(\forall x \in D)(f(x) \in \text{NULLS} \wedge g(x) \in \text{NULLS})$, for any $f$ and $g$; trivially, this is incoherent, as it would mean that both $f$ and $g$ take only null values.

Check through propositional calculus: $f \,\neg|\!-\, g \wedge \neg\neg|\!-\, f \cdot g$ can be computed as $(f \vee g) \wedge (f \text{ XNOR } g)$, which is equal to $f \wedge g$, i.e., $(\forall x \in D)(f(x) \in \text{NULLS} \wedge g(x) \in \text{NULLS})$, which is absurd.

v. (both $\neg f \,|\!-\, g$ and $\neg f \,\neg|\!-\, g$)

Suppose by *reductio ad absurdum* that $\neg f \,|\!-\, g \wedge \neg f \,\neg|\!-\, g$, i.e., $(\forall x \in D)(f(x) \in \text{NULLS} \Rightarrow g(x) \notin \text{NULLS} \wedge f(x) \in \text{NULLS} \Rightarrow g(x) \in \text{NULLS})$, for any $f$ and $g$; trivially, this is incoherent, as $g(x)$ may not simultaneously take both null and not null values.

Check through propositional calculus: $\neg f \,|\!-\, g \wedge \neg f \,\neg|\!-\, g$ can be computed as $(\neg f \vee \neg g) \wedge (\neg f \vee g)$, which is equal to $\neg f$, i.e., $(\forall x \in D)(f(x) \notin \text{NULLS})$, which is absurd (as $f$ is not totally defined).

vi. (both $\neg f \,|\!-\, g$ and $\neg\neg|\!-\, f \cdot g$)

Suppose by *reductio ad absurdum* that $\neg f \,|\!-\, g \wedge \neg\neg|\!-\, f \cdot g$, i.e., $(\forall x \in D)(f(x) \in \text{NULLS} \Rightarrow g(x) \notin \text{NULLS} \wedge (f(x) \in \text{NULLS} \wedge g(x) \in \text{NULLS} \vee f(x) \notin \text{NULLS} \wedge g(x) \notin \text{NULLS}))$, for any $f$ and $g$; trivially, this is incoherent, as it reduces to $(\forall x \in D)(f(x) \notin \text{NULLS} \wedge g(x) \notin \text{NULLS})$.

Check through propositional calculus: $\neg f \,|\!-\, g \wedge \neg\neg|\!-\, f \cdot g$ can be computed as $(\neg f \vee \neg g) \wedge (f \text{ XNOR } g)$, which is equal to $\neg f \wedge \neg g$, i.e., $(\forall x \in D)(f(x) \notin \text{NULLS} \wedge g(x) \notin \text{NULLS})$, which is absurd (neither $f$, nor $g$ are totally defined). *Q.E.D.*

*Corollary* 8: MatBase must always reject any of the 6 pairs from Proposition 8.

Please note that the pair $< f \,\neg|\!-\, g, \neg f \,|\!-\, g >$ can be computed as $(f \vee g) \wedge (\neg f \vee \neg g)$, which is equal to $f$ XOR $g$ ($f \oplus g$), i.e., $(\forall x \in D)(f(x) \in \text{NULLS} \oplus g(x) \in \text{NULLS})$, i.e., for any $x$ in $D$, either $f(x) \in$ NULLS or $g(x) \in$ NULLS but never both, which is theoretically legitimate, but not for any $f$ and $g$: e.g., if we consider $f$ = *SSN* and $g$ = *ITIN*, as shown above, *SSN* ¬|– *ITIN*; however, ¬*SSN* |– *ITIN* does not hold, as there are U.S. residents who do not have either *SSN* or *ITIN* (e.g., recently arrived / undocumented, dependents who do not file taxes, persons without tax filling obligations, and exempt religious groups).

Let us now analyze all the C(5, 3) = 10 triples made of the independent 5 existence constraint subtypes.

*Proposition* 9. No triple made of the independent 5 existence constraint subtypes over same two functions ($f$ and $g$) may exist in any coherent and minimal set of existence constraint subtypes.

*Proof*:

i. $<f \mid\!- g, f \neg\!\mid\!- g, \neg f \mid\!- g>$ is incoherent, according to Proposition 8 (i) and (ii).
ii. $<f \mid\!- g, f \neg\!\mid\!- g, \neg f \neg\!\mid\!- g>$ is incoherent, according to Proposition 8 (i) and (v).
iii. $<f \mid\!- g, f \neg\!\mid\!- g, \neg\neg\!\mid\!- f \cdot g>$ is incoherent, according to Proposition 8 (i) and (iv).
iv. $<f \mid\!- g, \neg f \mid\!- g, \neg f \neg\!\mid\!- g>$ is incoherent, according to Proposition 8 (ii) and (v).
v. $<f \mid\!- g, \neg f \mid\!- g, \neg\neg\!\mid\!- f \cdot g>$ is incoherent, according to Proposition 8 (ii) and (vi).
vi. According to Proposition 7 (i), $<f \mid\!- g, \neg f \neg\!\mid\!- g, \neg\neg\!\mid\!- f \cdot g> = <\neg\neg\!\mid\!- f \cdot g, \neg\neg\!\mid\!- f \cdot g>$, pair not minimal, as it contains duplicates, which is rejected according to Proposition 3.
vii. $<f \neg\!\mid\!- g, \neg f \mid\!- g, \neg f \neg\!\mid\!- g>$ is incoherent, according to Proposition 8 (v).
viii. $<f \neg\!\mid\!- g, \neg f \mid\!- g, \neg\neg\!\mid\!- f \cdot g>$ is incoherent, according to Proposition 8 (vi).
ix. $<f \neg\!\mid\!- g, \neg f \neg\!\mid\!- g, \neg\neg\!\mid\!- f \cdot g>$ is incoherent, according to Proposition 8 (iii).
x. $<\neg f \mid\!- g, \neg f \neg\!\mid\!- g, \neg\neg\!\mid\!- f \cdot g>$ is incoherent, according to Proposition 8 (v). *Q.E.D.*

*Corollary* 9.

i. No quadruple or quintuple made of the independent 5 existence constraint subtypes over same two functions (*f* and *g*) may exist in any coherent and minimal set of existence constraints.
ii. *MatBase* must not accept any triple discussed in Proposition 9.

*Proof*: trivial, left to the reader (*hint*: no triples are coherent, according to Proposition 8).

***The (E)MDM meta-model for the (E)MDM existence constraint subtypes***

Figure 5 shows the sets and functions from the (E)MDM meta-model of (E)MDM (which is also the (E)MDM model of the corresponding fragment of *MatBase's* metacatalog) that are referenced in the meta-model for the (E)MDM existence constraint subtypes shown in Figure 6 (in (E)MDM, Boolean functions need not be declared as totally defined, as null values are treated as *False*).

***SETS*** (The collection of all managed sets)

***FUNCTIONS*** (The set of all managed functions)
*#F* : *FUNCTIONS* ↔ NAT(maxNAT) total
*Domain* : *FUNCTIONS* → *SETS* total
*Total* : *FUNCTIONS* → BOOLE
**arity = card*({*#FP* | *Product*(*#FP*) = *#F*})

***FUNCT_PRODUCTS*** (The set of pairs <*pf*, *mf*> storing the member functions *mf* making the product function *pf*)
*#FP* : *FUNCT_PRODUCTS* ↔ NAT(maxNAT) total
*Product* : *FUNCT_PRODUCTS* → *FUNCTIONS* total
*FunctionMember* : *FUNCT_PRODUCTS* → *FUNCTIONS* total

***CONSTRAINTSET*** (The set of all managed constraints)
*#C* : *CONSTRAINTSET* ↔ NAT(maxNAT) total
*Set* : *CONSTRAINTSET* → *SETS*
*Mapping* : *CONSTRAINTSET* → *FUNCTIONS*

***Figure 5:*** The sets and functions referenced in the set, function, and constraint definitions from Figure 6.

***EXIST_CNSTRS* ⊆ *CONSTRAINTSET*** (The set of managed existence, non-existence, anti-existence, and inexistence constraints)

*Inexist?* : *EXIST_CNSTRS* → BOOLE

*Negation?* : *EXIST_CNSTRS* → BOOLE

*#C* : *EXIST_CNSTRS* ↔ *CONSTRAINTSET* total (canonical injection associated with ***EXIST_CNSTRS* ⊆ *CONSTRAINTSET***)

*ECLeftSide* : *EXIST_CNSTRS* → *FUNCTIONS*

*ECRightSide* : *EXIST_CNSTRS* → *FUNCTIONS* total

**Set = Domain° ECRightSide*

$C_0$: *ECLeftSide* ≠ *ECRightSide*

$C_1$: *ECRightSide* • *ECLeftSide* • *Inexist?* • *Negation?* key (There may not be two existence constraints having same right- and left-sides, and same values for *Inexist?* and *Negation?*)

$C_2$: ¬(*Inexist?* • *Negation?*) |— *ECLeftSide* (Any not negated non-inexistence constraint is an existence one, hence *ECLeftSide* is compulsory.)

$C_3$: ¬(*Inexist?* • *ECLeftSide*) |— *Negation?* (any non-inexistence constraint without left-side is a non-existence one, hence *Negation?* must be *True*.)

$C_4$: (∀*x*∈*EXIST_CNSTRS*)(*ECLeftSide*(*x*) ∉ NULLS ⇒ *Domain*(*ECLeftSide*(*x*)) = *Domain*(*ECRightSide*(*x*)) = *Set*(*#C*(*x*))) (Whenever the left-side of a(n) (non-)existence constraint is specified, then the domains of its both sides must be equal and equal to the set to which the constraint is associated.)

$C_5$: (∀*x*∈*EXIST_CNSTRS*)(*ECLeftSide*(*x*) ∈ NULLS ⇒ *ECRightSide*(*x*) = *Mapping*(*#C*(*x*))) (Whenever the left-side of a non-existence constraint is not specified, then the constraint is associated to its right-side.)

$C_6$: (∀*x*∈*EXIST_CNSTRS*)(*ECLeftSide*(*x*) ∈ NULLS ⇒ *Negation?*(*x*) ∧ **arity*(*ECRightSide*(*x*)) > 1) (Whenever the left-side of a non-existence constraint is not specified, then its right-side must be a function product and *Negation?* must be *True*.)

$C_7$: (∀*x*∈*EXIST_CNSTRS*)((*ECLeftSide*(*x*) ∉ NULLS ⇒ ¬*Total*(*ECLeftSide*(*x*))) ∧ ¬*Total*(*ECRightSide*(*x*))) (The right-side of a(n) (non-)existence constraint must not be totally defined and, whenever its left-side is specified, then it must not be totally defined either.)

$C_8$: (∀*x*∈*EXIST_CNSTRS*)(*ECLeftSide*(*x*) ∉ NULLS ∧ **arity*(*ECLeftSide*(*x*)) > 1 ⇒ ((∀*y*∈*FUNCT_PRODUCTS*)(*Product*(*y*) = *Product*(*ECLeftSide*(*x*)) ⇒ ¬*Total*(*FunctionMember*(*y*))))) (Whenever the left-side of a(n) (non-)existence constraint is specified and is a function product, then none of its members may be totally defined.)

$C_9$: (∀*x*∈*EXIST_CNSTRS*)(**arity*(*ECRightSide*(*x*)) > 1 ⇒ ((∀*y*∈*FUNCT_PRODUCTS*)(*Product*(*y*) = *Product*(*ECRightSide*(*x*)) ⇒ ¬*Total*(*FunctionMember*(*y*))))) (Whenever the right-side of a(n) (non-)existence constraint is a function product, then none of its members may be totally defined.)

$C_{10}$: (∀*x*∈*EXIST_CNSTRS*)(*ECLeftSide*(*x*) ∉ NULLS ∧ **arity*(*ECLeftSide*(*x*)) > 1 ∧ **arity*(*ECRightSide*(*x*)) = 1 ⇒ ((∀*y*∈*FUNCT_PRODUCTS*)(*Product*(*y*) = *Product*(*ECLeftSide*(*x*)) ⇒ *FunctionMember*(*y*) ≠ *ECRightSide*(*x*))) (Whenever the left-side of a(n) (non-)existence constraint is specified and is a function product, while the right-side is not a product, then none of its members may be equal to the right side.)

$C_{11}$: (∀*x*∈*EXIST_CNSTRS*)(*ECLeftSide*(*x*) ∉ NULLS ∧ **arity*(*ECLeftSide*(*x*)) > 1 ∧ **arity*(*ECRightSide*(*x*)) > 1 ⇒ ((∀*y*,*z*∈*FUNCT_PRODUCTS*)(*Product*(*y*) = *Product*(*ECLeftSide*(*x*)) ∧ *Product*(*z*) = *Product*(*ECRightSide*(*x*)) ⇒ *FunctionMember*(*y*) ≠ *FunctionMember*(*z*))) (Whenever the left-side of a(n) (non-)existence constraint is specified and is a function product, just like the right-side, then none of its members may be equal to any right-side members, i.e., no single function may be present on both sides of a(n) (non-)existence constraint.)

***Figure 6:*** The (E)MDM meta-model for the (E)MDM existence constraint subtypes.

### *MatBase pseudocode algorithms for enforcing the (E)MDM existence constraints*

Figures 7 to 12 show the *MatBase* SQL-embedded pseudocode algorithms for enforcing satisfiability, coherence, and minimality of (E)MDM existence constraint subtype sets.

### *Enforcing in- and anti-existence constraints in MatBase*

Enforcement of existence and non-existence constraints in *MatBase* was described in [1] (see Figures 2 to 4). Obviously, just like in their case, deleting an element from the domain set *D* underlying an anti-existence or inexistence constraint may never violate it, while, dually, inserting a new one or updating an existing one may. Consequently, to enforce such constraints, the class *D* of the software app that manages the corresponding db must contain an event-driven method *BeforeUpdate* like the one shown in Figure 2 from [1]. This method must be automatically launched whenever users ask (implicitly or explicitly) for saving either a new added line or the updates to *f* and/or *g* values on an existing line from table *D* (uniquely identified by the value of *D*'s surrogate primary key *x*). *MatBase* automatically generates it the first time that a constraint involving *D* is added to the db scheme.

```
Algorithm ECEA: Satisfiability, Coherence, and Minimality of (Anti/Non/In/)Existence Constraint Sets
Input: the current constraint set C, the required operation type ot (true for add or false for delete), constraint name
    cn, Boolean Negation?, Boolean Inexist?, left-side function f = f1 • … • fn : Df → Cf, and right-side function
    g = g1 • … • gm : Dg → Cg, n, m naturals;
Output: C — {cn}, if cn's deletion is confirmed, or C ∪ {cn}, if cn is admissible, or C otherwise;
Strategy (enforces Corollaries 0 to 9):
if ¬ot then if cn ∈ C then if user confirms his/her request to delete cn then C = C — {cn};
                              delete from the f • g's class the code for enforcing cn; end if;
            else display "Request rejected: " & cn & " is not a known constraint name!"; end if;
else if there is a constraint in C named cn then
       display "Request rejected: " & cn & " is the name " & "of another constraint of this db! Please choose " &
                                                                        "a unique constraint name instead!";
     else if there is a constraint cn' ∈ C having same values for f, Negation?, Inexist?, and g then
            display "Request rejected: this constraint is already stored in this db schema under the name " & cn';
          else if m == 0 then display "Request rejected: any existence constraint must have a right-side!";
               else if n == 0 ∧ m < 2 then display "Request rejected: as the left-side is null, the right-side must be"
                                                                                    & " a function product!";
                    else if there is no set D such that Df ⊆ D and Dg ⊆ D then display "Request rejected : " & f &
                                                            " and " & g & " do not have compatible domains!";
                         else Boolean ok = checkExistTotality(f, g);
```

```
                              if ok ∧ ¬Inexist? ∧ ¬Negation? ∧ n == 0 then display "Request rejected: the left-side"
                                                                      & " of an existence constraint is mandatory!"
                              else if ok ∧ ¬Inexist? ∧ n > 0 ∧ ¬Negation? then display "Request rejected: " &
                                                                  "Negation? is mandatory for non-existence constraints!"
                                   else if ok ∧ n > 0 ∧ m > 0 then ok = checkExistDisjoint(f, g);
                                        if ok then ok = checkExistCoherence(cn, Inexist?, f, Negation?, g);
                                           if ok then ok = checkExistSatisf(cn, Inexist?, f, Negation?, g);
                                              if ok then ok = checkExistRedundancies(cn, Inexist?, f,
                                                                                            Negation?, g);
                                                 if ok then C = C ∪ {cn}; add to D's class the
                                                                              code for enforcing cn;
                                                 end if;
                                              end if;
                                           end if;
                                        end if;
                                   end if;
                              end if;
                         end if;
                    end if;
               end if;
          end if;
     end if;
end if;
End Algorithm ECEA;
```

***Figure 7:*** *MatBase ECEA* Algorithm for enforcing satisfiability, coherence, and minimality of existence constraint sets.

Similar to the case of existence and non-existence ones, whenever *MatBase* accepts a single inexistence constraint *iec*: $\neg f$ |– $g$ over $D$ or a consolidated (compacted) one |– $f \cdot g$, it adds to this method a line that reads "if not Cancel then Cancel = *enforce_inexist_cnstr*(*iec*, f(x), g(x));"; whenever it accepts a single anti-existence constraint *aec*: $\neg f$ ¬|– $g$ over $D$ or a consolidated one ¬¬|– $f \cdot g$, it adds a line that reads "if not Cancel then Cancel = *enforce_anti-exist_cnstr*(aec, f(x), g(x));". These lines implement the $C = C \cup \{cn\}$ statement of the *Algorithm ECEA* from Figure 7.

Dually, whenever *MatBase* deletes an inexistence constraint named *iec* over $D$, it removes from this method the line that starts with "if not Cancel then Cancel = *enforce_inexist_cnstr*(*iec*,"; whenever it deletes an anti-existence constraint named *aec* over $D$, it removes the line that starts with "if not Cancel then Cancel = *enforce_anti-exist_cnstr*(aec,": this is how *MatBase* implements the $C = C - \{cn\}$ statement of the *Algorithm ECEA* from Figure 7.

```
Boolean Function checkExistTotality(function f = f1 • … • fn, function g = g1 • … • gm);
Input: two functions of an existence constraint of any of the 7 subtypes from Figure 4 defined on a set D
Returns: True, if all their member functions are partially defined, or False otherwise.
Strategy (enforces Corollary 1(iii)):
checkExistTotality = True;
DECLARE @STR VARCHAR(1000);
if n == 1 then SELECT @STR = FunctionName FROM FUNCTIONS
                  WHERE FunctionName = 'f' AND Total = 1;
if m == 1 then SELECT @STR = ISNULL(@STR + ',','') + FunctionName FROM FUNCTIONS
                                   WHERE FunctionName = 'g' AND Total = 1;
if n > 1 ∨ m > 1 then
  SELECT DISTINCT @STR = ISNULL(@STR + ',','') + M_FUNCTIONS.FunctionName
  FROM (SETS JOIN (FUNCTIONS JOIN FUNCT_PRODUCTS
          ON FUNCTIONS.[#F] = FUNCT_PRODUCTS.Product)
          ON SETS.[#S] = FUNCTIONS.Domain) JOIN FUNCTIONS AS M_FUNCTIONS
          ON FUNCT_PRODUCTS.FunctionMember = M_FUNCTIONS.[#F]
  WHERE FUNCTIONS.FunctionName IN ('f', 'g') AND SetName = 'D' AND M_FUNCTIONS.Total = 1;
if len(@STR) > 0 then checkExistTotality = False;
                  display "Request rejected: member function(s) " & @STR & " is/are totally defined!";
end if;
End Function checkExistTotality;
```

***Figure 8:*** Function *checkExistTotality* of *MatBase ECEA* Algorithm.

Figures 13 and 14 present the pseudocode of methods *enforce_inexist_cnstr* and *enforce_anti-exist_cnstr*, respectively, that *MatBase* is storing in its library *Constraints* [3], alongside the *enforce_existence_cnstr* and *enforce_non-existence_cnstr* shown in Figures 3 and 4 from [1].

## Results and Discussion

*Proposition* 10. (Characterization of Algorithm ECEA from Figures 7 to 12).

*Algorithm ECEA* from Figures 7 to 12 has the following properties:

i. It never loops infinitely.
ii. It is extremely fast, having complexity $O(k)$, $k$ natural.
iii. It enforces all 10 above Corollaries 0 to 9.
iv. It warrants the satisfiability, coherence, and minimality of the sets of all 7 subtypes from Figure 4.
v. It is sound.
vi. It is complete.
vii. It is optimal.

*Proof*:

i. Trivial, as it has no loops.
ii. Obvious, as in the worst case (i.e., when a new constraint is added to the existing set) the algorithm from Figure 7 performs $k_{max}$ = 15 steps; similarly, the function *checkExistTotality* from Figure 8 performs at most 7 steps, *checkExistDisjoint* from Figure 9 at most other 7, *checkExistCoherence* from Figure 10 another 2, *checkExistSatisf* from Figure 11 another 5, and *checkExistRedundancies* from Figure 12 another 6; in total, $k_{max}$ = 15 + 27 = 42.
iii. *checkExistTotality* from Figure 8 enforces Corollary 1, *checkExistDisjoint* from Figure 9 enforces Corollary 2, *checkExistCoherence* from Figure 10 Corollaries 8 and 9, and *checkExistRedundancies* from Figure 12 Corollaries 6 and 7, while *ECEA* from Figure 7 enforces Corollaries 0, 3, 4, 5.

```
Boolean Function checkExistDisjoint(function f = f1 • … • fn, function g = g1 • … • gm);
Input: two (Cartesian product) functions f and g defined on a same set D
Returns: False, if there is a fi that is among the gjs, or True otherwise.
Strategy (enforces Corollary 2(i)):
checkExistDisjoint = True;
nvarchar @fmname = null;
if n == 1 ∧ m == 1 then if f == g then checkExistDisjoint = False; display "Request rejected: " & f & " = " & g &
                    "!"; end if;
elseif n == 1 ∧ m > 1 then
  SET @fmname = SELECT FUNCTIONS_1.FunctionName
  FROM ((SETS JOIN FUNCTIONS ON SETS.[#S] = FUNCTIONS.Domain) JOIN FUNCT_PRODUCTS
     ON FUNCTIONS.[#F] = FUNCT_PRODUCTS.Product) JOIN FUNCTIONS AS FUNCTIONS_1
     ON FUNCT_PRODUCTS.FunctionMember = FUNCTIONS_1.[#F]
  WHERE FUNCTIONS.FunctionName = "g" AND SetName = "D" AND FUNCTIONS_1.FunctionName = "f"
  ;
elseif n > 1 ∧ m = 1 then
  SET @fmname = SELECT FUNCTIONS_1.FunctionName
  FROM ((SETS JOIN FUNCTIONS ON SETS.[#S] = FUNCTIONS.Domain) JOIN FUNCT_PRODUCTS
     ON FUNCTIONS.[#F] = FUNCT_PRODUCTS.Product) JOIN FUNCTIONS AS FUNCTIONS_1
     ON FUNCT_PRODUCTS.FunctionMember = FUNCTIONS_1.[#F]
  WHERE FUNCTIONS.FunctionName = "f" AND SetName = "D" AND FUNCTIONS_1.FunctionName = "g"
  ;
else
 SET @fmname = SELECT Max(MEMBER_FUNCTIONS.FunctionName)
 FROM ((FUNCT_PRODUCTS AS FUNCT_PRODUCTS_1
     JOIN (SETS JOIN (
          (FUNCTIONS JOIN FUNCT_PRODUCTS ON FUNCTIONS.[#F] = FUNCT_PRODUCTS.Product)
        JOIN FUNCTIONS AS MEMBER_FUNCTIONS
        ON FUNCT_PRODUCTS.FunctionMember = MEMBER_FUNCTIONS.[#F]
        ) ON SETS.[#S] = FUNCTIONS.Domain
       ) ON FUNCT_PRODUCTS_1.FunctionMember = MEMBER_FUNCTIONS.[#F]
    ) JOIN FUNCTIONS AS FUNCTIONS_1 ON FUNCT_PRODUCTS_1.Product = FUNCTIONS_1.[#F])
    JOIN SETS AS SETS_1 ON FUNCTIONS_1.Domain = SETS_1.[#S]
 WHERE FUNCTIONS.FunctionName "f" AND SETS.SetName = "D" AND FUNCTIONS_1.FunctionName =
       "g" AND SETS_1.SetName = "D";
end if;
if @fmname ≠ null then checkExistDisjoint = False;
   display "Request rejected: " & @fmname & " is a member function of both " & f & " and " & g & "!";
end if;
End Function checkExistDisjoint;
```

***Figure 9:*** Function *checkExistDisjoint* of *MatBase ECEA* Algorithm.

```
Boolean Function checkExistCoherence(string cn, Boole Inexist?, function f, Boole Negation?, function g);
Input: an existence constraint of any of the 7 subtypes from Figure 4
Returns: False, if adding cn to the current constraint set C would make C incoherent, or True otherwise.
Strategy (enforces Corollaries 8 and 9):
checkExistCoherence = True;
if ¬Inexist? ∧ ¬Negation? ∧ ((C includes a constraint cn': f ¬|— g) ∨ (C includes a constraint cn': ¬f |— g)) ∨
  ¬Inexist? ∧ Negation? ∧ ((C includes a constraint cn': ¬f ¬|— g) ∨ (C includes a constraint cn': ¬¬ |— f •g))
  ∨ Inexist? ∧ ¬Negation? ∧ ((C includes a constraint cn': ¬f |— g) ∨ (C includes a constraint cn': ¬¬ |— f •
  g)) then checkExistCoherence = False; display "Request rejected: constraint set would become incoherent!";
end if;
End Function checkExistCoherence;
```

***Figure 10:*** Function checkExistCoherence of MatBase ECEA Algorithm.

```
Boolean Function checkExistSatisf(string cn, Boole Inexist?, function f, Boole Negation?, function g);
Input: an existence constraint of any of the 7 subtypes from Figure 4 defined on a set D
Returns: True, if cn is satisfied by the current db instance, or False otherwise.
Strategy:
checkExistSatisf = True; int @x = null;
select case cn's subtype
  case f |— g: SET @x = SELECT Max(x) FROM D WHERE f IS NOT NULL AND g IS NULL;
  case f ¬|— g:
  case ¬|— f • g: SET @x = SELECT Max(x) FROM D WHERE f IS NOT NULL AND g IS NOT NULL;
  case ¬f |— g:
  case |— f • g: SET @x = SELECT Max(x) FROM D WHERE f IS NULL AND g IS NULL;
  case ¬f ¬|— g: SET @x = SELECT Max(x) FROM D WHERE f IS NULL AND g IS NOT NULL;
  case ¬¬|— f • g: SET @x = SELECT Max(x) FROM D WHERE f IS NULL AND g IS NOT NULL OR
                                                      f IS NOT NULL AND g IS NULL;
  else: checkExistSatisf = False; display cn & " is not well-formed!"
end select;
if @x ≠ null then checkExistSatisf = False; display "Request rejected: " & @x & " violates " & cn & "!";
End Function checkExistSatisf;
```

***Figure 11:*** Function *checkExistSatisf* of *MatBase ECEA* Algorithm.

iv. Function *checkExistSatisf* from Figure 11 guarantees satisfiability; function *checkExistCoherence* from Figure 10 coherence, and function *checkExistRedundancies* from Figure 12, as well as *ECEA*, which is enforcing Corollary 3, minimality.

v. Soundness is guaranteed by Corollaries 0, 1, 2, 4, and 5, all of them enforced by *ECEA* and its functions *checkExistTotality* and *checkExistDisjoint*: as it accepts only well-formed constraints, its output will always contain only valid sets of constraints of all 7 subtypes from Figure 4.

vi. Completeness is also guaranteed by Corollaries 0 to 2 and 4 to 5: any well-formed constraint of all 7 subtypes from Figure 4 are accepted.

vii. Optimality is obvious from inspecting the algorithms from Figures 7 to 12: all of them are performing their tasks with the minimum number possible of instructions and disk accesses, be it to the *MatBase* metacatalog tables or to the underlying db instance. *Q.E.D.*

*Proposition* 11. (Characterization of method *enforce_inexist_cnstr* from Figure 13).

Method *enforce_inexist_cnstr* from Figure 13 has the following properties:

i. It never loops infinitely.
ii. It is fast, having complexity $O(n + m)$, $n$, $m$ naturals being the arities of the left- and right-side of the corresponding constraint, respectively.
iii. It correctly enforces any stored inexistent constraint, be it simple or consolidated.
iv. It is sound.
v. It is complete.
vi. It is optimal.

```
Boolean Function checkExistRedundancies(string cn, Boole Inexist?, function f, Boole Negation?, function g);
Input: an existence constraint of any of the 7 subtypes from Figure 4
Returns: False, if adding cn to the current constraint set C would make C redundant, or True otherwise.
Strategy (enforces Corollaries 6 and 7):
checkExistRedundancies = False;
if ¬Inexist? ∧ Negation? ∧ n > 0 ∧ (C includes a constraint cn': ¬├─ f • g) then
  display "Request rejected: this constraint is implied by enforced constraint " & cn' & "!";
else if ¬Inexist? ∧ Negation? ∧ n = 0 ∧ g == h • i ∧ (C includes a constraint cn': h ¬├─ i) then replace cn' by
      cn; display "Enforced constraint " & cn': h ¬├─ i & " is replaced by " & cn & " for minimality reasons!";
   else if n == 0 ∧ g = h • i ∧ Inexist? ∧ Negation? then // request to add constraint of subtype ¬¬├─ f • g
      if C includes a constraint cn': ¬h ¬├─ i then replace cn' by cn; display "Constraint " & cn' &
                                      "'s body was replaced by " & cn & "'s one for minimality reasons!";
      else if C includes constraint cn': h ├─ i then
         if C includes constraint cn'': ¬h ¬├─ i then replace cn' and cn'' by cn; display "Constraints "
               & cn' & " and " & cn'' & " were replaced by " & cn & " for minimality reasons!";
         else replace cn' by cn;
            display "Constraint " & cn' & " was replaced by " & cn & " for minimality reasons!";
         end if;
        end if;
      end if;
     else if C includes a constraint cn': ¬¬├─ f • g then
        if ¬Inexist? ∧ ¬Negation? ∨ Inexist? ∧ Negation? then
         display "Request rejected: enforced constraint " & cn' & " implies " & cn & "!";
       else if ¬Inexist? ∧ ¬Negation? ∧ (C includes a constraint cn': ¬f ¬├─ g) ∨
           Inexist? ∧ Negation? ∧ (C includes a constraint cn': f ├─ g) then replace body of cn' with
           ¬¬├─ f • g; display "Constraint accepted. Consequently, " & cn & " and enforced constr"
          & "aint " & cn' & "'s body were replaced by " & ¬¬├─ f • g & " for minimality reasons!";
         end if;
       end if;
     end if;
   end if; checkExistRedundancies = True;
end if;
End Function checkExistRedundancies;
```

***Figure 12:*** Function *checkExistRedundancies* of *MatBase ECEA* Algorithm.

*Proof*:

i. Trivial, as it has 3 loops, out of which 2 are executing at most $m$ times and one at most $n$ times.
ii. Obvious, as in the worst case (i.e., when the constraint is a single one and the new values for $f(x)$ and $g(x)$ are valid) the algorithm performs $k_{max} = 7 + n + m$ steps.
iii. For consolidated ones, in its first loop it checks whether at least one ($f \cdot g$)'s function member (in this case consolidated as the third method parameter) has a not null value for the current element $x$ of the underlying domain; for single ones, it first checks in the second loop whether there is at least one $f$'s function member for which the corresponding current $x$ desired value is null and then, in the third loop, whether all $g$'s function members have not null values for the current $x$ element. For both subtypes of inexistent constraints, whenever the corresponding conditions are not met, the method rejects saving the new values for $f(x)$ and $g(x)$ with appropriate error messages.
iv. Soundness is guaranteed by (iii) above, as this method rejects any attempt to violate any inexistence constraint, keeping the db instances clean of invalid data.
v. Completeness is guaranteed by the fact that it accepts any inexistence constraint and possible values for its functions $f$ and $g$ (accepting only those that satisfy the constraint and rejecting those that would violate it).
vi. Optimality is obvious from inspecting the algorithm from Figure 13: this method is performing its tasks with the minimum number possible of instructions and disk accesses, be it to the *MatBase* metacatalog tables or to the underlying db instance, for both single and consolidated inexistence constraints. *Q.E.D.*

```
Boolean method enforce_inexist_cnstr(iec, f(x) = (f1 • … • fn)(x), g(x) = (g1 • … • gm)(x));
// returns True if the values of f(x) and g(x) violate iec and False otherwise
enforce_inexist_cnstr = False;
if x is a new line or f(x) has been modified or g(x) has been modified then
  int i = 1; Boolean null = True;
  if f(x) == ∅ then          // consolidated set of inexistence constraints
    while ¬enforce_inexist_cnstr ∧ i ≤ m ∧ null
        if gi(x) ∉ NULLS then null = False; else i = i + 1; end if;
    end while;
    if null then enforce_inexist_cnstr  = True;
              display "Saving these values is rejected: according to inexistence constraint "
                   & iec & ", at least a column of " & g & " must have a not null value!";
    end if;
  else                       // single inexistence constraint
    null = False;
    while ¬null ∧ i ≤ n
        if fi(x) ∈ NULLS then null = True; else i = i + 1; end if;
    end while;
    if null then i = 1;
      while null ∧ i ≤ m
        if gi(x) ∈ NULLS then null = False; else i = i + 1; end if;
      end while;
      if ¬null then enforce_inexist_cnstr = True;
        display "Saving these values is rejected: according to inexistence constraint " &
                     iec & ", all columns of " & g & " must have not null values!"
      end if;
    end if;
  end if;
end if;
End method enforce_inexist_cnstr;
```

***Figure 13:*** *MatBase* method *enforce_inexist_cnstr*.

*Proposition* 12. (Characterization of method *enforce_anti-exist_cnstr* from Figure 14).

Method *enforce_anti-exist_cnstr* from Figure 14 has the following properties:

i. It never loops infinitely.
ii. It is fast, having complexity $O(n + m)$, $n$, $m$ naturals being the arities of the left- and right-side of the corresponding constraint, respectively.
iii. It correctly enforces any stored anti-existent constraint, be it simple or consolidated.
iv. It is sound.
v. It is complete.
vi. It is optimal.

```
Boolean method enforce_anti-exist_cnstr(aec, f(x) = (f1 • … • fn)(x), g(x) = (g1 • … • gm)(x));
// returns True if the values of f(x) and g(x) violate aec and False otherwise
enforce_anti-exist_cnstr = False;
if x is a new line or f(x) has been modified or g(x) has been modified then int i = 2; Boolean null;
  if f(x) == ∅ then          // consolidated set of anti-existence constraints
    if g1(x) ∈ NULLS then null = True; else null = False; end if;
    while ¬enforce_anti-exist_cnstr ∧ i ≤ m
        if gi(x) ∈ NULLS ∧ null ∨ gi(x) ∉ NULLS ∧ ¬null then i = i + 1;
        else enforce_anti-exist_cnstr = True;
             display "Saving these values is rejected: according to anti-existence constraint "
                & aec & ", all columns of " & g & " must have either null or not null values!";
        end if;
    end while;
  else                       // single anti-existence constraint
    null = True; i = 1;
    while null ∧ i ≤ n
        if fi(x) ∉ NULLS then null = False; else i = i + 1; end if;
    end while;
    if null then i = 1;
      while null ∧ i ≤ m
        if gi(x) ∉ NULLS then null = False; else i = i + 1; end if;
      end while;
      if ¬null then enforce_anti-exist_cnstr = True;
        display "Saving these values is rejected: according to anti-existence constraint " &
                        aec & ", all columns of " & g & " must have null values!"
      end if;
    end if;
  end if;
end if;
End method enforce_anti-exist_cnstr;
```

***Figure 14:*** *MatBase* method *enforce_anti-exist_cnstr*.

*Proof*:

i. Trivial, as it has 3 loops, out of which 2 are executing at most $m$ times and one at most $n$ times.
ii. Obvious, as in the worst case (i.e., when the constraint is a single one and the new values for $f(x)$ and $g(x)$ are valid) the algorithm performs $k_{max} = 7 + n + m$ steps.
iii. For consolidated ones, in its first loop it checks whether all $(f \cdot g)$'s function members (in this case consolidated as the third method parameter) have either null or not null values for the current element $x$ of the underlying domain; for single ones, it first checks in the second loop whether there is at least one $f$'s function member for which the corresponding current $x$ desired value is not null and then, in the third loop, whether all $g$'s function members have null values for the current $x$ element. For both sub-types of anti-existent constraints, whenever the corresponding conditions are not met, the method rejects saving the new values for $f(x)$ and $g(x)$ with appropriate error messages.
iv. Soundness is guaranteed by (*iii*) above, as this method rejects any attempt to violate any anti-existence constraint, keeping the db instances clean of invalid data.
v. Completeness is guaranteed by the fact that it accepts any anti-existence constraint and possible values for its functions $f$ and $g$ (accepting only those that satisfy the constraint and rejecting those that would violate it).
vi. Optimality is obvious from inspecting the algorithm from Figure 14: this method is performing its tasks with the minimum number possible of instructions and disk accesses, be it to the *MatBase* metacatalog tables or to the underlying db instance, for both single and consolidated anti-existence constraints. *Q.E.D.*

For example, let us consider *D* = *PERSONS*, *C* = {*iec*: ¬|– *email* · *PhoneNo*, *aec*: ¬*PassedAwayYear* ¬|– *KilledBy*}, *iec'*: ¬|– *email* · *Name*, with *Name* totally defined, *iec*": ¬*email* |– *email* · *PhoneNo*, *aec'*: ¬¬|– *PassedAwayYear* · *KilledBy*, and *aec*'': ¬*PassedAwayYear* |– *KilledBy*;

1. Obviously, both *iec* and *aec* are accepted by Algorithm *ECEA*, so that the *BeforeUpdate* method of class *PERSONS* would look as in Figure 15.
2. Any attempt to save to the db a line containing {(*email* · *PhoneNo*)(*x*) = <,>} is rejected with the error message "Saving these values is rejected: according to inexistence constraint *iec*, at least a column of *email* · *PhoneNo* must have a not null value!".
3. Any attempt to save to the db a line containing {(*PassedAwayYear*(*x*), *KilledBy*(*x*) = < , 123>} is rejected with the error message "Saving these values is rejected: according to anti-existence constraint *aec*, all columns of *PassedAwayYear* · *KilledBy* must have a null value!".
4. Requests to save to the db lines containing {(*email* · *PhoneNo*)(*x*) = <"christian.mancas@gmail.com",>}, or {(*email* · *PhoneNo*)(*x*) = <"someone@gmail.com", 12345678>}, or {(*email* · *PhoneNo*)(*x*) = <, 12345678>} are accepted.
5. Requests to save to the db lines containing {(*PassedAwayYear*(*x*), *KilledBy*(*x*)) = <1610, 123>} or {(*PassedAwayYear*(*x*), *KilledBy*(*x*)) = < , >} are accepted.
6. Request to add *iec'* to *C* is rejected with the error message "Request rejected: member function(s) Name is/are totally defined!".
7. Request to add *iec*" to *C* is rejected with the error message "Request rejected: email is member function Name of both email and email · PhoneNo!".
8. Request to add *aec'* to *C* is granted but with the info message "Enforced constraint *aec*: ¬*PassedAwayYear* ¬|– KilledBy is replaced by *aec'* for minimality reasons!". Automatically, *MatBase* replaces in the method *BeforeUpdate* from Figure 15 the line reading "if not Cancel then Cancel = *enforce_anti-exist_cnstr*(*aec*, *PassedAwayYear*(*x*), *KilledBy*(*x*));" with the following one: "if not Cancel then Cancel = *enforce_anti-exist_cnstr*(*aec*, , (*PassedAwayYear* · *KilledBy*)(*x*))".
9. Request to add *aec*" to *C* is rejected with the error message "Request rejected: constraint set would become incoherent!".
10. Request to delete *iec* from *C* would trigger *MatBase* to display the message "Are you sure you want to delete constraint *iec*: ¬|– *email* · *PhoneNo*? Cancel OK"; if user chooses Cancel, nothing happens; if he/she chooses OK, then *MatBase* deletes from the method *BeforeUpdate* from Figure 15 the line reading "if not Cancel then Cancel = *enforce_inexist_cnstr*(*iec*, , (*email* · *PhoneNo*)(*x*));".

In the C# and SQL Server version of *MatBase*, the method *BeforeUpdate* from Figure 15 is implemented as a similar one attached to the event type *Validating*. Moreover, *MatBase* is also enforcing all non-relational constraints in T-SQL, thus securing the data integrity at the db level as well, protecting its instance from direct user db access. For example, Figure 16 shows a fragment of the *MatBase* generated T-SQL trigger associated to table *PERSONS* for enforcing constraint *iec*.

```
method BeforeUpdate();
// enforces constraints on table PERSONS whenever more than one column value is updated
Boolean Cancel = False;
if not Cancel then Cancel = enforce_inexist_cnstr(iec, , (email • PhoneNo)(x));
if not Cancel then Cancel = enforce_anti-exist_cnstr(aec, PassedAwayYear(x), KilledBy(x));
if Cancel then reject saving the current line into the db;
end method BeforeUpdate;
```

***Figure 15:*** The event-driven method *BeforeUpdate* of class *PERSONS* associated to table *PERSONS*.

```
1   USE [PersonsDB]
2   GO
3
4   /****** Object:  Trigger [dbo].[TR_Persons_Validate] Script Date: 5/19/2026 10:30:29 PM ******/
5   SET ANSI_NULLS ON
6   GO
7
8   SET QUOTED_IDENTIFIER ON
9   GO
10
11  CREATE   TRIGGER [dbo].[TR_Persons_Validate]
12  ON [dbo].[PERSONS]
13  AFTER INSERT, UPDATE
14  AS
15  BEGIN
16      SET NOCOUNT ON;
17      SET XACT_ABORT ON;
18
19      -- ════════════════════════════════════════════
20      -- PART 1 – PERSONS non-relational constraints
21      -- ════════════════════════════════════════════
22
23      -- ── Constraint iec: not|- email • PhoneNo ──────────────
24      IF EXISTS (
25          SELECT 1 FROM inserted i
26          WHERE  i.email IS NULL
27            AND  i.PhoneNo IS NULL
28      )
29      BEGIN
30          RAISERROR('Constraint iec violated: Either email or/and PhoneNo must be not null.', 16, 1);
31          ROLLBACK TRANSACTION; RETURN;
32      END
```

***Figure 16:*** Fragment of the *MatBase* generated T-SQL trigger associated to table *PERSONS* for enforcing constraint *iec*.

## Conclusions

Pushed by examples encountered in conceptual data modeling of several subuniverses of interest, we added to our (Elementary) Mathematical Data Model four new constraint types, dual to the existence and non-existence ones: the single and consolidated inexistence and anti-existence ones.

We defined them formally, established their well-forming rules, and fully axiomatized the set of all 7 subtypes of existence constraints, by establishing and proving the propositions and corollaries that govern the satisfiability, coherence, and minimality of such constraint type sets.

Based on them, we updated the *MatBase* algorithm managing the existence and non-existence constraints to also deal with inexistence and anti-existence ones. We proved that this algorithm has constant complexity, being extremely fast, and is guaranteeing the satisfiability, coherence, and minimality of existence constraint sets containing all their 7 subtypes, while being sound, complete, and optimal.

We also designed and implemented in *MatBase* two SQL-embedded pseudocode algorithms for enforcing inexistence and anti-existence constraints, respectively, for both single and consolidated ones. We proved that they have linear complexity in the sum of the arities of their involved functions and that they are sound, complete, and optimal, while correctly enforcing any such constraint.

We provided real-life examples of constraints of all four new subtypes introduced and shown that they are correctly handled by the enforcement algorithms.

While the new mathematical concepts and results added by this research are not outstanding, their applicability to the conceptual data modeling theory and database software applications design and development is noteworthy: currently, to enforce the business rules in the provided examples, as well as in any other similar ones, developers must design, develop, and test dedicated code for each of them, having no clue whatsoever that they belong to a certain constraint subtype, having certain properties. Consequently, they are not only prone to design and programming errors, but they also risk blocking users from storing valid data (whenever the corresponding constraint set becomes incoherent) or to slow down db and app response time (whenever the corresponding constraint set becomes not minimal, i.e., includes redundant constraints). Obviously, even if developers do not have access at a *MatBase* copy, they can still use the algorithms described and characterized in this paper.

Consequently, we consider that this paper contributes significantly to the database constraint theory and practice, in one of the subdomains that is difficult to deal with: correctly managing the null values.

***Conflict of interest***

The authors declare that the research was conducted in the absence of any commercial or financial relationships that could be construed as a potential conflict of interest.

***Acknowledgements***

This research was not sponsored by anybody and nobody other than its author contributed to it. The author is grateful to Diana Christina Mancas, who implemented in *MatBase* and tested the algorithms presented in this paper, and to Mihaela Virginia Mancas, who is always carefully checking all our manuscripts.